\documentclass[aps,10pt,pra,twocolumn,showpacs,amsmath,amssymb, superscriptaddress]{revtex4-1}

\usepackage{graphicx}
\usepackage{color}
\usepackage{transparent}
\usepackage{hyperref}
\usepackage{txfonts}
\usepackage{amsmath}
\usepackage{braket}

\begin{document}

\title{Quantum effects in amplitude death of coupled anharmonic self-oscillators}

\author{Ehud Amitai}
\affiliation{Department of Physics, University of Basel, Klingelbergstrasse 82, 4056 Basel, Switzerland}
\author{Martin Koppenh\"{o}fer}
\affiliation{Department of Physics, University of Basel, Klingelbergstrasse 82, 4056 Basel, Switzerland}
\author{Niels L\"{o}rch}
\affiliation{Department of Physics, University of Basel, Klingelbergstrasse 82, 4056 Basel, Switzerland}
\author{Christoph Bruder}
\affiliation{Department of Physics, University of Basel, Klingelbergstrasse 82, 4056 Basel, Switzerland}

\begin{abstract}
Coupling two or more self-oscillating systems may stabilize their
zero-amplitude rest-state, therefore quenching their oscillation. This
phenomenon is termed ``amplitude death". Well-known and studied in
classical self-oscillators, amplitude death was only recently
investigated in quantum self-oscillators [Ishibashi et al.,~Phys. Rev. E \textbf{96}, 052210 (2017)]. Quantitative differences between the classical and
quantum descriptions were found. Here, we demonstrate that for quantum
self-oscillators with anharmonicity in their energy spectrum, multiple
resonances in the mean phonon number can be observed. This is a result
of the discrete energy spectrum of these oscillators, and is not
present in the corresponding classical model. Experiments can be
realized with current technology and would demonstrate these genuine quantum effects in the amplitude death phenomenon.  
\end{abstract}
\maketitle

\section{Introduction}
Self-sustained oscillators are a class of oscillating systems, in which the amplitude of the periodic motion is maintained by an incoherent power source, which is balanced by a nonlinear energy loss \cite{Pikovsky, Balanov}. The phase of the self-oscillator is therefore not fixed by the phase of the power source. This phase freedom allows the self-oscillator to lock or entrain its phase to the phase of an external signal, or to the phase of another self-oscillator, a phenomenon known as synchronization \cite{Pikovsky, Balanov}. 

The synchronization of quantum oscillators has become a very active research topic in recent years, due to advances in experiments with micro- and nanomechanical oscillators \cite{Lee, Walter, Weiss, Lee2, Walter2, Ameri, Bastidas, Loerch2, Sonar, Amitai, Loerch, Xu, Weiss2016}. The quantum van der Pol (vdP) oscillator was proposed as a generic model for a quantum self-oscillator \cite{Walter, Lee}, allowing for the investigation of synchronization in the quantum regime. Synchronization of a quantum vdP oscillator to a drive \cite{Walter, Lee, Weiss}, the synchronization of two mutually coupled vdP oscillators \cite{Lee, Lee2, Walter2, Ameri}, and the synchronization of networks of such oscillators \cite{Bastidas}, were theoretically investigated. Also, it was shown that using a squeezing Hamiltonian instead of a harmonic drive can produce stronger synchronization \cite{Sonar}. Recently, genuine quantum effects in the synchronization of such vdP oscillators were predicted \cite{Loerch2}.

The quantum vdP oscillator model, being a quantum model for a self-oscillator, can be used to study other phenomena, different than quantum synchronization. 
Still, much less effort has been invested in that direction. Recently,
the quantum amplitude dynamics of two dissipatively coupled quantum
vdP oscillators has been studied \cite{Ishibashi}. In the classical
case, it is known that dissipatively coupling two self-oscillators may
stabilize their zero-amplitude rest-state via a Hopf bifurcation
\cite{Aronson, Ermentrout, Mirollo1}. This phenomenon, in which the
amplitude of the two self-oscillators is strongly suppressed as they
approach their steady state, is known in the literature as ``amplitude
death" or ``oscillation death" \cite{Pikovsky, Saxena, Koseska,
  Koseska2}. While both terms are often used, Ref.~\cite{Koseska2}
distinguishes the case in which both oscillators approach an identical
steady state, and the case in which each oscillator approaches a
different steady state. ``Amplitude death" refers to the former, while
``oscillation death" refers to the latter. We keep this nomenclature,
and use the term ``amplitude death" for the case described in our paper. In Ref.~\cite{Ishibashi}, the researchers have shown the quantum-analog of the amplitude death phenomenon. They have found quantitative differences when comparing the quantum model with a corresponding classical model with Gaussian noise.  

Here, we investigate the amplitude dynamics of two dissipatively coupled quantum vdP oscillators with anharmonicity in their energy spectrum. We report \emph{qualitative} differences in the amplitude death phenomenon between the quantum model and a corresponding classical model with Gaussian noise. For increasing detuning between the two oscillators, we observe a decay in the oscillation amplitude, as expected in amplitude death. Then however, for an even larger detuning, we observe an increase of the oscillation amplitude. We demonstrate that such an increase is the result of the quantized anharmonic energy spectrum. It is, to the best of our knowledge, the first time that genuine quantum features that go beyond a semiclassical drift-diffusion picture are predicted to exist in the amplitude death phenomenon. 

This paper is organized as follows. We describe the models used in this paper in Sec.~\ref{The_Model}. This includes the quantum model, the noiseless classical model, and the semiclassical model, i.e.~ a classical model with Gaussian noise. In Sec.~\ref{noise_induced_oscillation_death}, we describe the effect of the anharmonicity in the energy spectrum on the vdP oscillation amplitude. We show that this anharmonicity leads to strong oscillation-amplitude suppression, however only in the presence of noise. Genuine quantum effects in the amplitude death phenomenon, which are not the result of noise, but stemming from the quantized energy levels of the anharmonic oscillators, are described in Sec.~\ref{Quantum_effects}. In Sec.~\ref{Experimental_Realizations} we conclude and remark about possible experimental realizations of the proposed system. 

\begin{figure}[b] 
\includegraphics[width=\columnwidth]{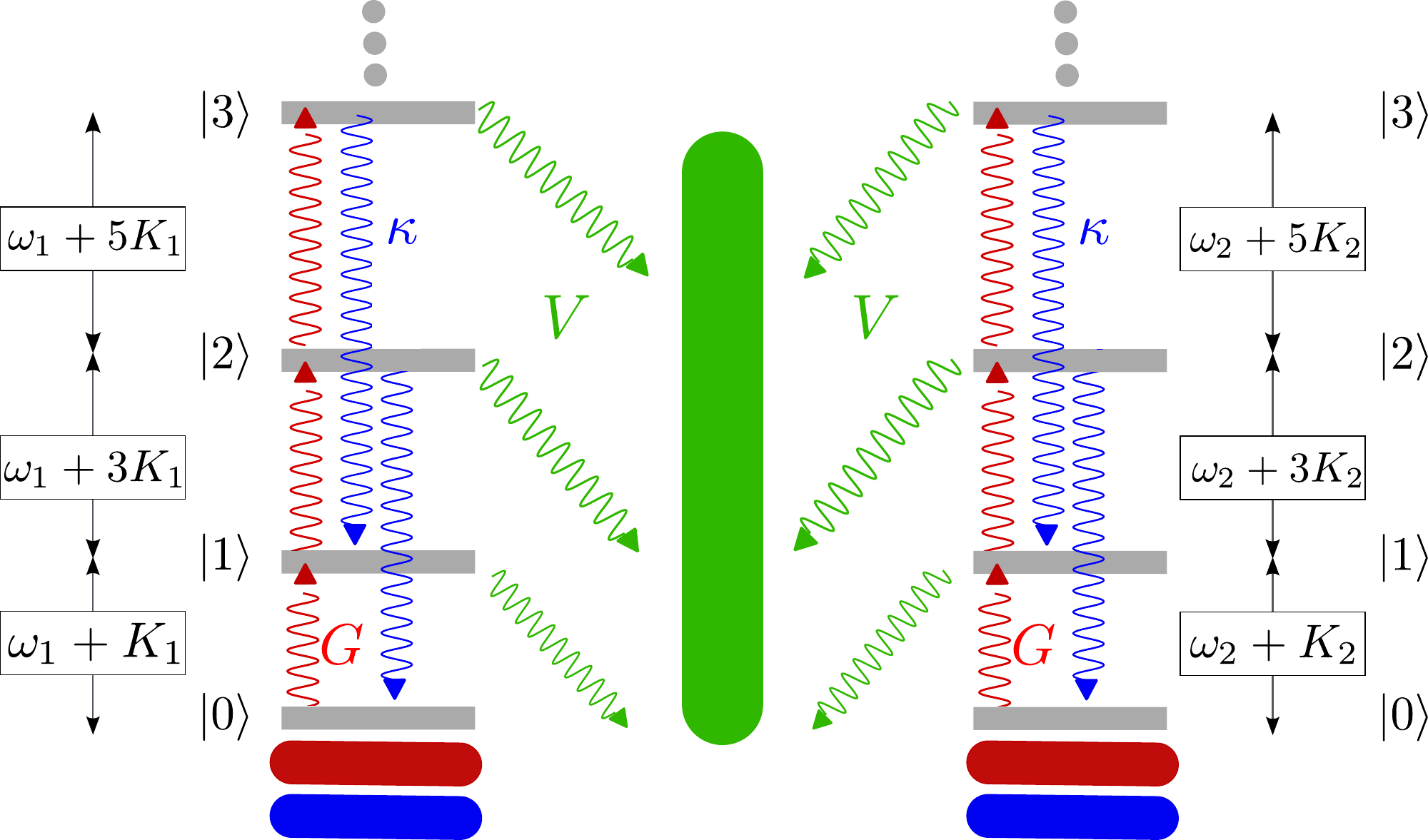}
\caption{The four lowest-lying discrete energy levels of the quantum vdP oscillator with Kerr nonlinearity. The Kerr nonlinearity leads to an energy level spacing $\omega_m + (2n+1)K_m$ between the $n$-th and $(n+1)$-th energy levels. The wiggly lines describe non-unitary processes stemming from coupling the system to Markovian reservoirs (marked by rectangles with rounded corners): The incoherent energy gain with rate $G$ and the incoherent nonlinear energy loss with rate $\kappa$ are obtained by coupling the individual vdP oscillators to their own Markovian reservoirs. The dissipative coupling with strength $V$ is obtained by coupling the vdP oscillators to a common Markovian reservoir.}
\label{Fig:system}
\end{figure}

\section{The Model}   \label{The_Model}
We consider two anharmonic dissipatively coupled quantum vdP oscillators
\cite{Ishibashi, Walter2, Lee2}. The schematics of the energy spectrum of the oscillators and the non-unitary processes involved in the coupling to the environment are shown in Fig.~\ref{Fig:system}. The time evolution of the density matrix $\rho$ of the two oscillators is governed by the quantum master equation $(\hbar=1)$
\begin{equation} \label{master_equation}
	\partial_t
	\rho
=
	\sum_{m=1}^2
	\left(
		-i[H_m, \rho]
	+
		G\mathcal{D}[a_m^\dagger]\rho
	+
		\kappa\mathcal{D}[a_m^2]\rho
	\right)
+
	V\mathcal{D}[a_1-a_2]\rho,
\end{equation}  
where $a_m$ and $a_m^\dagger$ are the annihilation and creation
operators of the $m$-th oscillator,  and $H_m = \omega_m a_m^\dagger
a_m + K_m(a_m^\dagger a_m)^2$ is the Hamiltonian of the $m$-th
oscillator, with $\omega_m$ and $K_m$ being the natural frequency and
the Kerr nonlinearity parameter of the $m$-th oscillator,
respectively. This Hamiltonian leads to an energy level spacing $\omega_m + (2n+1)K_m$ between the $n$-th and $(n+1)$-th energy levels of the $m$-th oscillator.
The non-unitary dynamics is described using Lindblad
operators, $\mathcal{D}[x] \rho \equiv x\rho x^\dagger -(x^\dagger x
\rho + \rho x^\dagger x)/2$. The parameters $G$ and $\kappa$ describe the rate of energy gain and the rate of nonlinear energy dissipation of the self-oscillators, respectively. $V$ defines the strength of the dissipative coupling. Such a dissipative coupling is obtained by assuming that the two vdP oscillators are coupled to a common Markovian reservoir \cite{Mogilevtsev}, as schematically shown in Fig.~\ref{Fig:system}. In the following, we will use QuTiP \cite{Johansson, Johansson2} to numerically simulate this master equation.    

In the model described by Eq.~(\ref{master_equation}), we have chosen $\kappa$ and $G$ to be identical for both vdP oscillators. This allows us to simplify our analysis by discarding any difference between the states of the two self-oscillators which may arise as a result of their individual character. This is by no means a crucial choice for observing the noise-induced amplitude death  and quantum effects described below. We have maintained the freedom of choosing a non-identical natural frequency, $\omega_m$, as the amplitude death depends critically on the frequency detuning between the two self-oscillator. Furthermore, we allow for non-identical Kerr nonlinearity, $K_m$, as it helps to elucidate the quantum effects described in Sec.~\ref{Quantum_effects}.

It is known that in the absence of a Kerr nonlinearity, the uncoupled ($V=0$) vdP oscillators exhibit limit-cycles. We would like to emphasize that this is also true in the presence of a Kerr nonlinearity. This is apparent when examining the steady state density matrix for such a Kerr nonlinear vdP oscillator, which is given by the diagonal $\rho^{(V=0)}_{nn}=(G/\kappa)^n \Phi(1+n, G/\kappa+n, G/\kappa)/\left[(G/\kappa)_n \Phi(1,G/\kappa,2G/\kappa)\right]$, where $\left(\cdot\right)_n$ denotes the Pochhammer symbol and $\Phi$ is Kummer's confluent hypergeometric function \cite{Dodonov1997, Loerch2}. $\rho^{(V=0)}$ depends only on $G/\kappa$ and not on the Kerr parameter $K_m$. It therefore describes limit-cycles with no preferred phase, just as for the harmonic $K_m=0$ case.

The equations of motion for the classical amplitudes of oscillation, $\alpha_m \equiv \braket{a_m}$, can be obtained from Eq.~(\ref{master_equation}). Using the Heisenberg equation of motion and after employing a mean-field approximation, one obtains 
\begin{equation} \label{EOM_classical}
	\partial_t
	\alpha_m
=
	-i\left[\omega_m +2K_m|\alpha_m|^2\right]\alpha_m
+
	\frac{G}{2} \alpha_m
-
	\kappa |\alpha_m|^2 \alpha_m
+
	\frac{V}{2} (\alpha_{\bar{m}}-\alpha_m),
\end{equation}  
for $m\in\left\{1,2\right\}$, and where $\bar{m}\neq m$. These equations of motion constitute our classical noiseless model. 

To obtain from Eq.~(\ref{master_equation}) a semiclassical model, i.e.~a classical model which includes Gaussian noise, we describe the system using a partial differential equation for the Wigner distribution function $W(\alpha_1, \alpha_1^*, \alpha_2, \alpha_2^*, t)$ \cite{Gardiner, Carmichael, Ishibashi},
 \begin{equation} \label{Full_EOM_W}
\begin{aligned}
	\partial_t
	W(\boldsymbol{\alpha})
=
	\sum_{m=1}^2&
	\left[
		-
		\left(
			\frac{\partial}{\partial \alpha_m}
			\mu_{\alpha_m}
		+
			\text{c.c.}
		\right)
\right.
\\
&
\left.
	+
		\frac{1}{2}
		\left(
			\frac{\partial^2}{\partial\alpha_m \partial \alpha_m^*}D_{\alpha_m \alpha_m^*}
		+
			\frac{\partial^2}{\partial\alpha_m \partial \alpha_{\bar{m}}^*}D_{\alpha_m \alpha_{\bar{m}}^*}
		\right)
\right.
\\
&
\left.
	+
		\frac{\kappa-iK_m}{4}
		\left(
			\frac{\partial^3}{\partial \alpha_m^* \partial \alpha^2_m}\alpha_m
		+
			\text{c.c.}
		\right)
	\right]
	W(\boldsymbol{\alpha}).
\end{aligned}
\end{equation}
This phase-space representation is completely equivalent to the master equation description, Eq.~(\ref{master_equation}). The drift coefficients are given by
\begin{equation}
\begin{aligned}
	\mu_{\alpha_m}
=&
	\left\{
		-i
		\left[
			\omega_m
		+
			2K_m |\alpha_m|^2
		\right]
	+
		\frac{G}{2}
	-
		\kappa
		\left(
			|\alpha_m|^2
		-
			1
		\right)
	-
		\frac{V}{2}
	\right\}
	\alpha_m
\\
&
+
	\frac{V}{2}\alpha_{\bar{m}},
\end{aligned}
\end{equation}
and the diffusion coefficients are given by
\begin{equation} \label{Diffusion}
\begin{aligned}
	D_{\alpha_m, \alpha_m^*} &= G + 2\kappa(2|\alpha_m|^2-1)+V,
\\
	D_{\alpha_m, \alpha_{\bar{m}}^*} &= -V.
\end{aligned}
\end{equation}

In the classical limit ($|\alpha_m|\gg 1$), we can neglect the third-order derivatives of Eq.~(\ref{Full_EOM_W}) \cite{Carmichael, Walls}. By doing so, we obtain the Fokker-Planck equation \cite{Risken}
\begin{equation} \label{FPE}
\begin{aligned}
	\partial_t
	W(\boldsymbol{\alpha})
=
	\sum_{m=1}^2&
	\left[
		-
		\left(
			\frac{\partial}{\partial \alpha_m}
			\mu_{\alpha_m}
		+
			\text{c.c.}
		\right)
\right.
\\
&
\left.
	+
		\frac{1}{2}
		\left(
			\frac{\partial^2}{\partial\alpha_m \partial \alpha_m^*}D_{\alpha_m \alpha_m^*}
		+
			\frac{\partial^2}{\partial\alpha_m \partial \alpha_{\bar{m}}^*}D_{\alpha_m \alpha_{\bar{m}}^*}
		\right)
	\right]
	W(\boldsymbol{\alpha})
.
\end{aligned}
\end{equation}
Eq.~(\ref{FPE}) constitutes our semiclassical model. It can be further transformed into an equivalent Langevin form \cite{Ishibashi}, which can be straightforwardly numerically simulated. The transformation is shown in Appendix~\ref{App:Langevin}.


\section{Noise-induced amplitude death}   \label{noise_induced_oscillation_death} 
\begin{figure}[t]   
\includegraphics[width=\columnwidth]{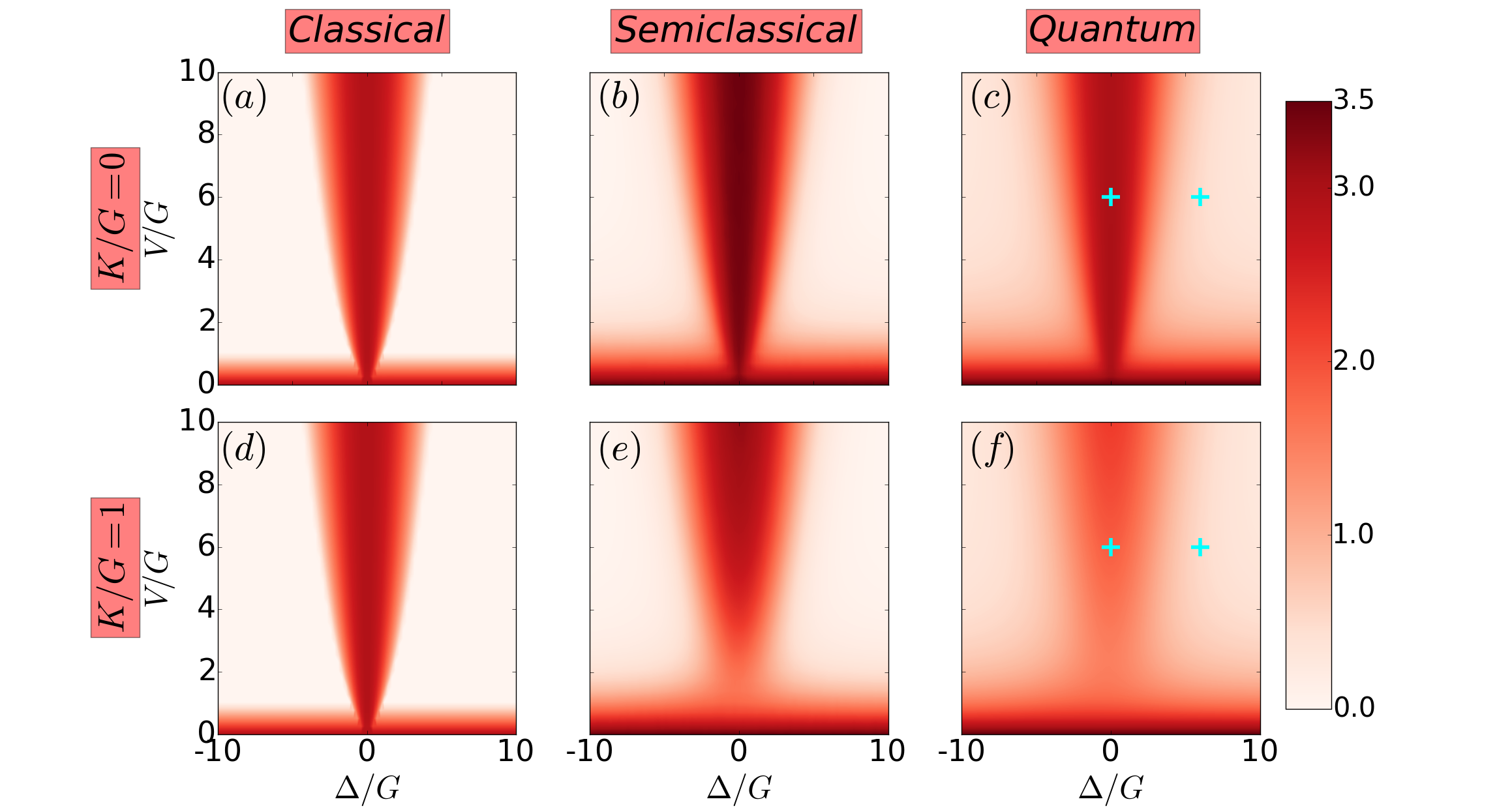}
\caption{Amplitude suppression of two coupled self-oscillators for the
  harmonic and anharmonic cases in the classical, semiclassical, and quantum
  descriptions. (a) and (d) show the squared amplitude $|\alpha_1|^2$ of the noiseless classical oscillator obtained from Eq.~(\ref{EOM_classical}). (b) and (e) present the long-time limit amplitude squared, $\overline{|\alpha_1|^2}$, obtained from numerically simulating the semiclassical model, Eq.~(\ref{FPE}), and then ensemble-averaging over many independent trajectories. (c) and (f) show the mean phonon number $\braket{a_1^\dagger a_1}$ of the quantum oscillator, Eq.~(\ref{master_equation}). The upper plots (a), (b), and (c) correspond to the harmonic case, $K/G=0$. The lower plots (c), (d) and (e) correspond to the anharmonic case, with $K/G=1$. A decrease in $\overline{|\alpha_1|^2}$ and $\braket{a_1^\dagger a_1}$ is seen in the anharmonic case, as compared with the harmonic case. The energy loss rate is $\kappa/G=0.2$ for all plots. Cyan crosses mark the parameters for which the Wigner density functions appearing in Fig.~\ref{Fig:Wigner} were calculated. }
\label{Fig:harmonic_amplitudes}
\end{figure}

The rest-state of two \emph{harmonic} self-oscillators is always unstable without a coupling between the two oscillators. When the two self-oscillators are dissipatively coupled, the rest-state may become stable, leading to strong amplitude suppression. This depends on the strength of the coupling $V$, and on the frequency detuning between the two self-oscillators, $\Delta\equiv \omega_2 - \omega_1$. In the classical noiseless case, it is predicted that the rest-state is stable in the regime $G<V<(\Delta^2 + G^2)/(2G)$ \cite{Aronson}. This behavior can be seen in Fig.~\ref{Fig:harmonic_amplitudes}~(a), which shows the squared amplitude $|\alpha_1|^2=|\alpha_2|^2$. For two vdP oscillators with an \emph{anharmonic} energy spectrum, the effective oscillation frequency of the individual oscillators, $\tilde{\omega}_m \equiv \omega_m + 2K_m|\alpha_m|^2$, depends on the amplitude of oscillation. This is a direct result of the anharmonicity in their energy spectrum, $K_m$. In the case that this anharmonicity is identical for both oscillators, $K_1=K_2=K$, the effective frequency detuning is identical to the natural frequency detuning,
\begin{equation}
	\tilde{\Delta}
=
	\tilde{\omega}_2 - \tilde{\omega}_1
=
	\omega_2 - \omega_1
+
	2K(|\alpha_2|^2-|\alpha_1|^2)
=
	\Delta,
\end{equation}
since the relation $\alpha_1 = \alpha_2$ holds in this case. We therefore expect the amplitude of oscillation of the oscillators with Kerr nonlinearity to be identical to the amplitude of oscillation of the harmonic oscillators, for any specific values of $V$ and $\Delta$. This is indeed the case, as can be seen by comparing Fig.~\ref{Fig:harmonic_amplitudes}~(a) with Fig.~\ref{Fig:harmonic_amplitudes}~(d), in which the squared amplitude of oscillation $|\alpha_1|^2=|\alpha_2|^2$ for $K/G=1$ is shown.  

\begin{figure}[t]   
\includegraphics[width=\columnwidth]{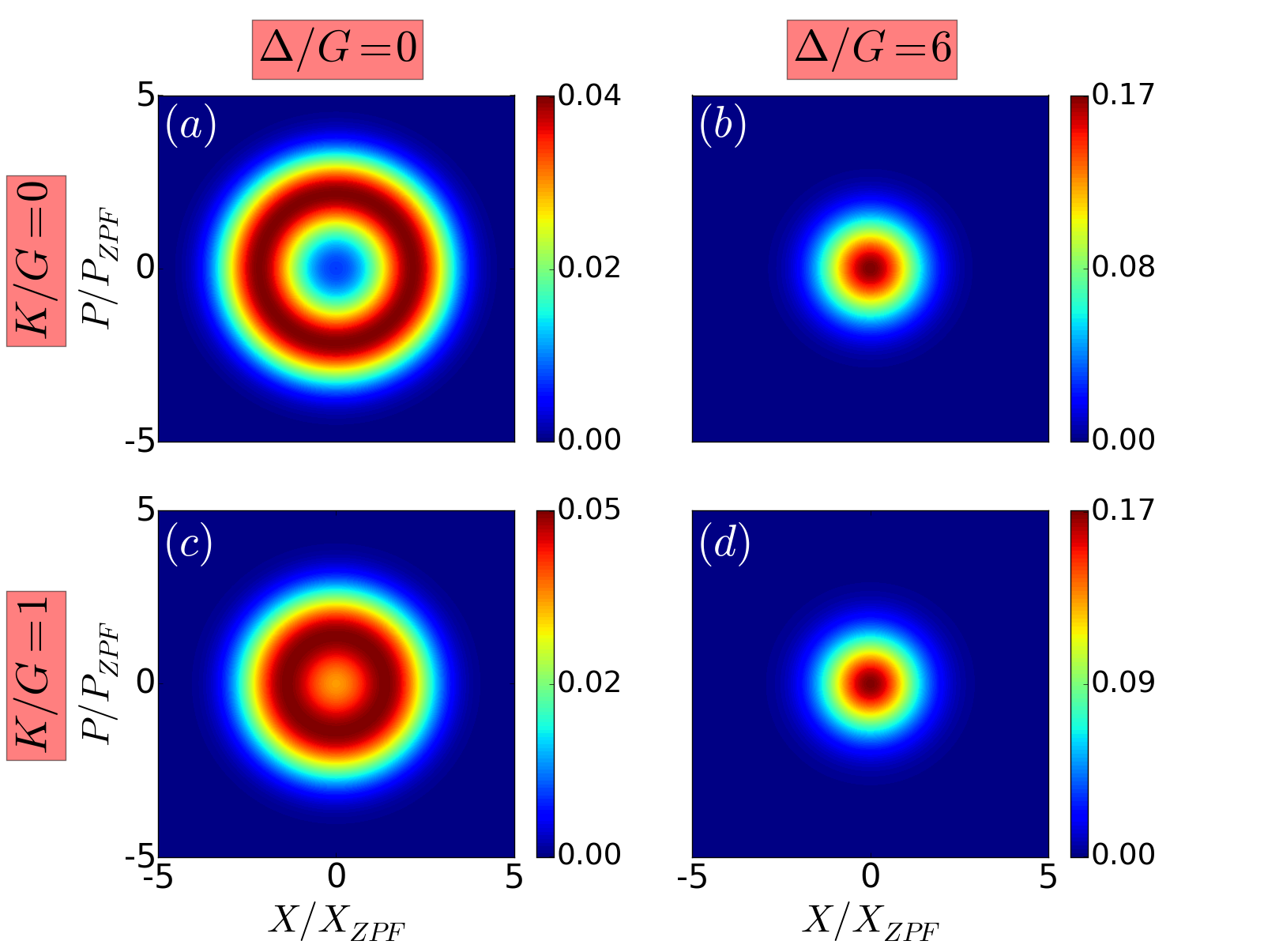}
\caption{Wigner density function for the steady state of the oscillator before and after amplitude death occurred, for both the harmonic case and the nonlinear case. (a) and (b) correspond to the cyan crosses marked in Fig.~\ref{Fig:harmonic_amplitudes}~(c), while (c) and (d) match the cyan crosses shown in Fig.~\ref{Fig:harmonic_amplitudes}~(f). The suppression of oscillation amplitude in the presence of a Kerr nonlinearity is clearly visible by comparing (a) and (c). The limit-cycle shrinks, resulting in a decrease of $\braket{a_1^\dagger a_1}$. The Wigner distributions are all rotationally symmetric, having no preferred phase.
In all plots, $\kappa/G=0.2$ and $V/G=6$.   }
\label{Fig:Wigner}
\end{figure}

For vdP oscillators in the presence of noise, on the other hand, the anharmonicity drastically changes the oscillation amplitude as compared with the harmonic case. This can be seen both in our semiclassical model, and in the fully quantum description. In Fig.~\ref{Fig:harmonic_amplitudes}~(b), we numerically simulate the semiclassical model, Eq.~(\ref{FPE}),  for $K=0$, and show the long-time limit amplitude squared, $\overline{|\alpha_1|^2}$, which is ensemble-averaged over many independent trajectories. This $\overline{|\alpha_1|^2}$ is shown as a function of both the detuning $\Delta$ and the coupling strength $V$. Oscillations are sustained for small enough $\Delta$, with slightly higher amplitudes than in the noiseless case. This oscillation amplitude is highly suppressed  in the regime where amplitude death is expected. Nevertheless, the amplitude of oscillation does not vanish completely, as noise hinders the complete collapse. This agrees with Ref.~\cite{Ishibashi}. In Fig.~\ref{Fig:harmonic_amplitudes}~(e), in which $\overline{|\alpha_1|^2}$ is shown for $K/G=1$, and in contrast to the classical noiseless case, the amplitudes of oscillation are significantly changed. This can be seen by comparing Fig.~\ref{Fig:harmonic_amplitudes}~(e) to Fig.~\ref{Fig:harmonic_amplitudes}~(b). It is seen that the values of $\overline{|\alpha_1|^2}$ for $V>G$ are significantly lower for the anharmonic case, as compared with the harmonic case.

As mentioned, a similar decrease is seen also in the quantum description. In Fig.~\ref{Fig:harmonic_amplitudes}~(c) we show the mean phonon number of the first oscillator $\braket{a^\dagger_1 a_1}$ for the harmonic case, as a function of $\Delta$ and $V$. As discussed in Ref.~\cite{Ishibashi}, the mean phonon number significantly decreases in the regime where amplitude death is expected classically, but does not vanish completely. Noise prevents the complete collapse. This can also be seen in Fig.~\ref{Fig:Wigner}~(a) and Fig.~\ref{Fig:Wigner}~(b), in which we plot the Wigner function representation of the steady state of the oscillator before and after amplitude death occurred. The parameters chosen for these Wigner representations are marked in cyan crosses in Fig.~\ref{Fig:harmonic_amplitudes}~(c). After amplitude death takes place, the probability distribution is sharply concentrated about the axis origin, leading to low phonon expectation values $\braket{a_1^\dagger a_1}$. When nonlinearity is introduced,  just as in the semiclassical description, the mean phonon number of the oscillators is significantly changed. This is seen in Fig.~\ref{Fig:harmonic_amplitudes}~(f), in which the mean phonon number $\braket{a^\dagger_1 a_1}$ is shown for $K/G=1$. As in the semiclassical description, it is seen that the values of $\braket{a^\dagger_1 a_1}$ for $V>G$ are significantly lower for the $K/G=1$ case, as compared with the $K/G=0$ case. This is also seen in the Wigner function representation, shown for the nonlinear case in Fig.~\ref{Fig:Wigner}~(c) and in Fig.~\ref{Fig:Wigner}~(d), for the parameters marked in cyan crosses in Fig.~\ref{Fig:harmonic_amplitudes}~(f). Even before amplitude death occurred, the limit-cycle of the oscillator shrank as compared with the harmonic case, Fig.~\ref{Fig:Wigner}~(a). The nonlinearity leads therefore to a decrease of $\braket{a_1^\dagger a_1}$ in the quantum case.
Note that while Fig.~\ref{Fig:harmonic_amplitudes} and Fig.~\ref{Fig:Wigner} show the average phonon number  and Wigner function representation of the first oscillator, almost identical figures are obtained for the second oscillator (see Fig.~\ref{Fig:Revival_Diff_combined}~(d) and discussion in the end of Sec.~\ref{Quantum_effects}).

The underlying cause for this decrease, seen in the semiclassical model and in the quantum description, is noise. When noise is present, the amplitude of the self-oscillator fluctuates, as is implied by the existence of a diffusion constant, Eq.~(\ref{Diffusion}). The effective frequency of the oscillators with Kerr nonlinearity, $\tilde{\omega}_m$, depends on this fluctuating amplitude of oscillation. For that reason, the frequency is now a fluctuating quantity as well. The bigger the anharmonicity $K$ is, the larger the frequency fluctuations become. This implies that when noise is present in the system, the spread of values for $\tilde{\Delta}$ is wider than the spread of values of the effective detuning for harmonic self-oscillators, $\Delta$. Therefore, increasing $K$ has a similar effect as increasing the effective detuning between the two self-oscillators. As the dissipative coupling is sensitive to the detuning, we see the effect of increasing $K$ as a decrease in $\braket{a_1^\dagger a_1}$ ($\braket{a_2^\dagger a_2}$) , for $V>G$. For $V\ll G$, on the other hand, the dissipative term plays only a minor role. Therefore, increasing $K$ does not significantly change the occupation number $\braket{a_1^\dagger a_1}$ ($\braket{a_2^\dagger a_2}$).

\begin{figure}[t]  
\includegraphics[width=\columnwidth]{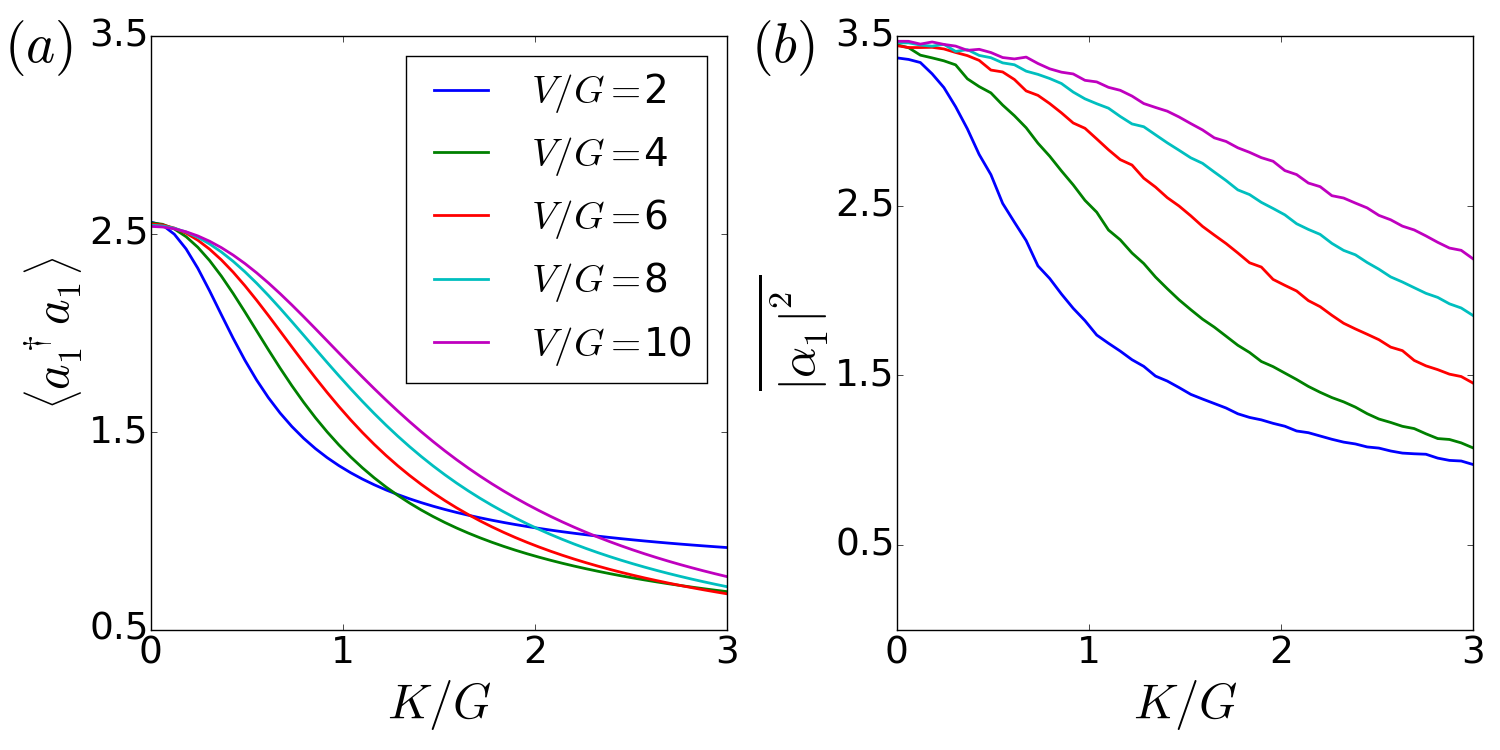}
\caption{(a) Average occupation number, $\braket{a_1^\dagger a_1}$, as a function of $K$. It is obtained using Eq.~(\ref{master_equation}). (b) Average oscillation amplitude squared obtained from Eq.~(\ref{FPE}), $\overline{|\alpha_1|^2}$, as a function of $K$. In both noisy models, a noticeable decrease in the oscillation amplitude is seen for increasing $K$. The legend shown describes both plots.   
Other parameters used in both plots are $(\kappa, \Delta)=(0.2, 0.0) \times G$. }
\label{Fig:n1(K)_combined}
\end{figure}

In Fig.~\ref{Fig:n1(K)_combined}~(a), we numerically simulate the quantum master equation, Eq.~(\ref{master_equation}), and show the decrease of $\braket{a^\dagger_1 a_1}$ for increasing $K$. In Fig.~\ref{Fig:n1(K)_combined}~(b), we numerically simulate the semiclassical model, Eq.~(\ref{FPE}), and show the decrease in the average amplitude squared, $\overline{|\alpha_1|^2}$, for increasing $K$. Indeed, both noisy models show this decrease, and only quantitative differences can be seen when comparing the two. The noiseless classical model cannot account for this amplitude suppression, as is seen in Fig.~\ref{Fig:harmonic_amplitudes}. We therefore conclude that this amplitude suppression, or average occupation number reduction, is noise-induced.

For very large values of $K/G$, this noise-induced amplitude
suppression can balance the amplitude growth induced by the linear
energy gain $G$. This allows us to set $\kappa=0$ for these cases,
while still keeping the self-oscillators in the quantum parameter
regime in which only a small number of energy levels is populated (see Sec.~\ref{Quantum_effects}). For smaller values of $K/G$, a finite $\kappa$ is required to keep the oscillators in the quantum parameter regime.

\section{Quantum effects - Amplitude revival}   \label{Quantum_effects}

In the  quantum parameter regime, the anharmonicity leads to genuine quantum effects in the amplitude death phenomenon, which cannot be modeled using a semiclassical model. They are the result of the quantized, discrete energy spectrum of the oscillators (see Fig.~\ref{Fig:system}). The Kerr anharmonicity $K_m$ leads to an energy level spacing $\omega_m + (2n+1)K_m$ between the $n$-th and the $(n+1)$-th Fock levels of the $m$-th anharmonic quantum vdP oscillator. There are therefore several discrete frequencies relevant for each oscillator. As the amplitude death phenomenon depends on the detuning between the frequencies of the oscillators, we can expect this discreteness to be reflected in the mean phonon number $\braket{a_m^\dagger a_m}$ of each oscillator. In order to observe this discreteness however, one must also consider the broadening of the energy levels due to the dissipative processes. Working in a parameter regime in which the energy level spacing is much larger than the energy level broadening is therefore a necessity.  

\begin{figure}[t] 
\begin{minipage}{\columnwidth}
	\includegraphics[width=\columnwidth]{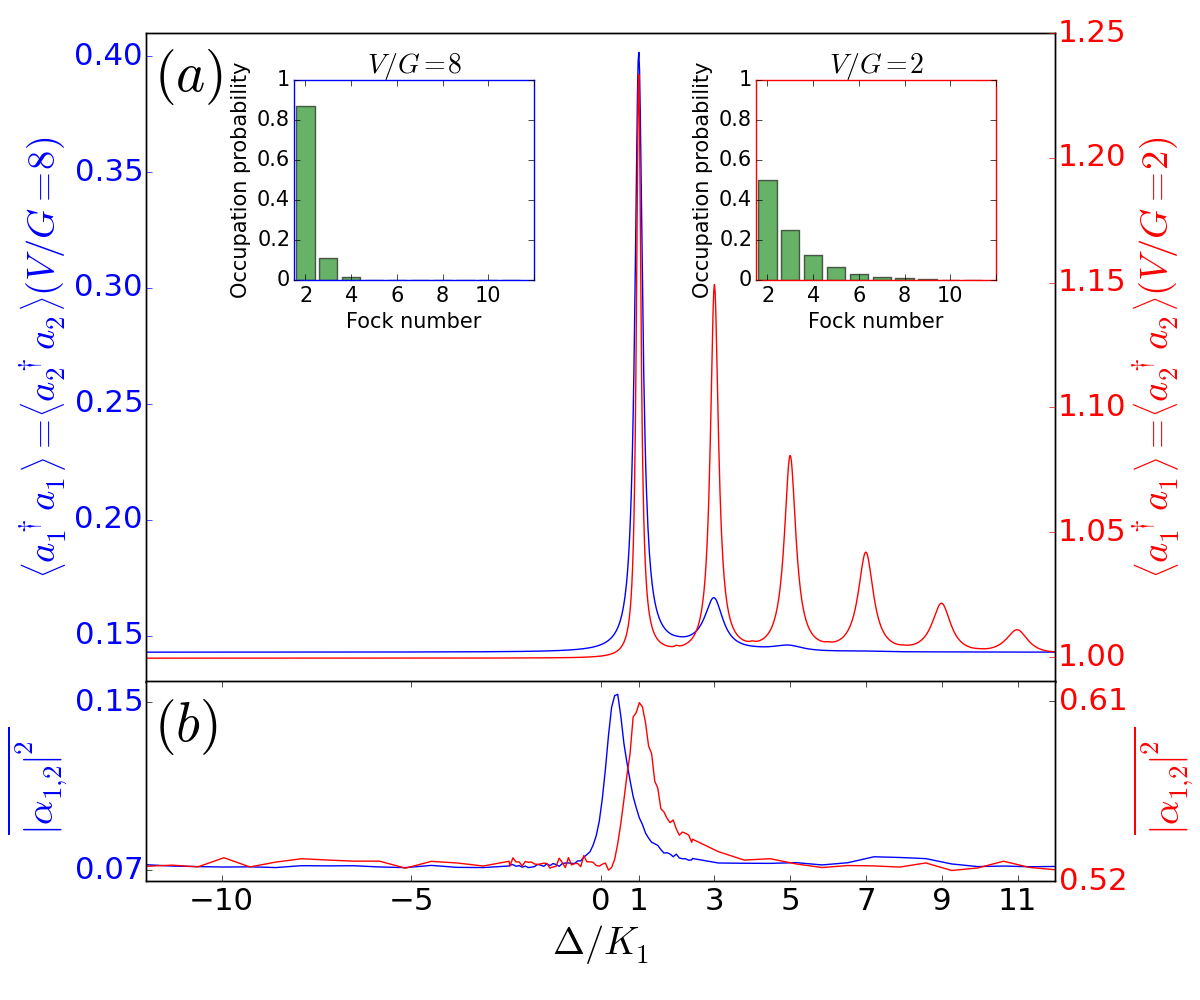}
\end{minipage}
\begin{minipage}{\columnwidth}
	\includegraphics[width=\columnwidth]{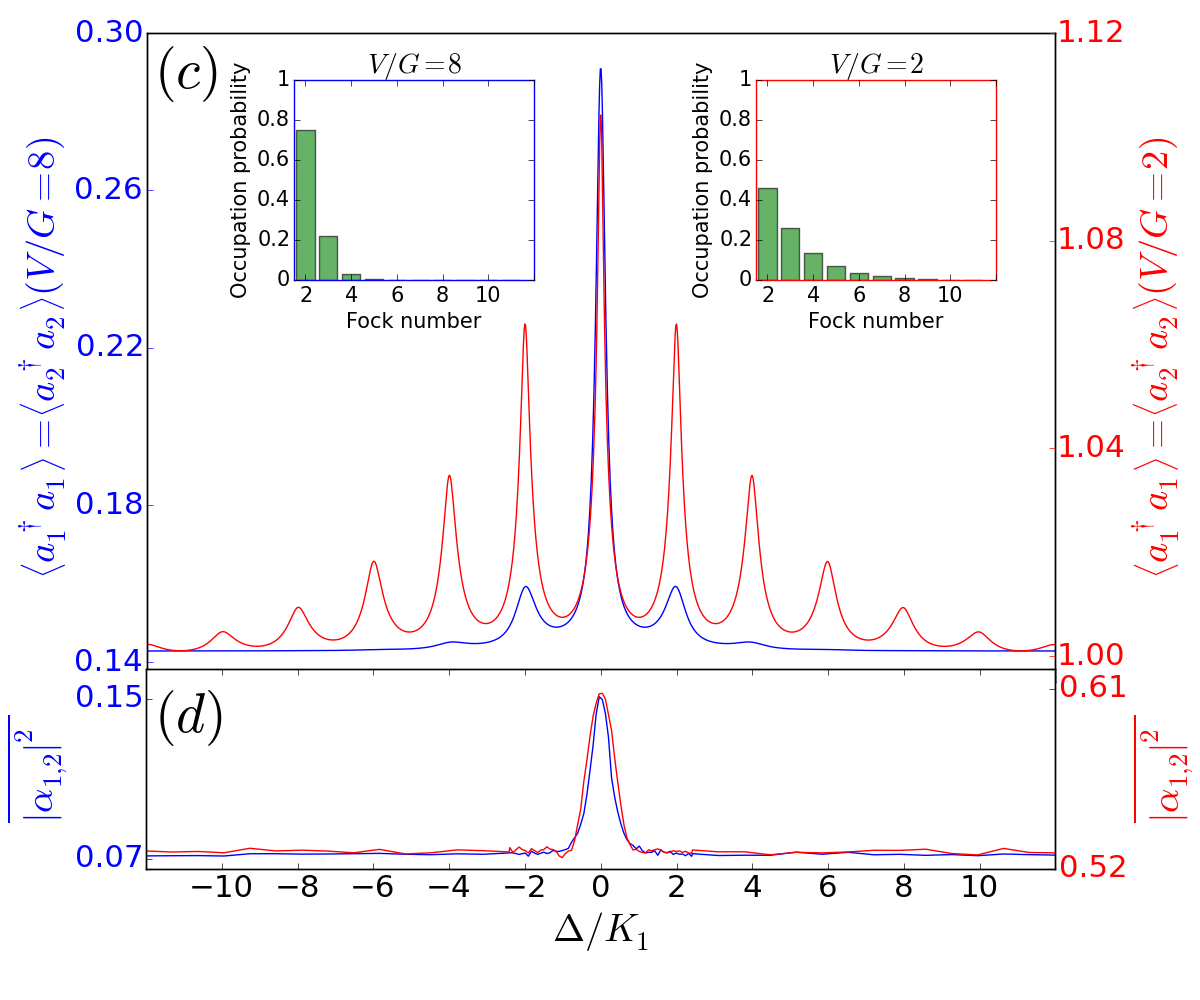}
\end{minipage}
\caption{Average occupation number $\braket{a_1^\dagger a_1}=\braket{a_2^\dagger a_2}$ and average oscillation amplitude squared $\overline{|\alpha_1|^2}=\overline{|\alpha_2|^2}$, obtained from Eq.~(\ref{master_equation}) and Eq.~(\ref{FPE}) respectively, are shown as a function of the detuning in (a) and (b) for a vdP oscillator with very strong nonlinearity $K_1/G=50$ coupled to a harmonic ($K_2/G=0$) vdP oscillator. In (c) and (d), the same quantities are shown for two anharmonic ($K_1/G=K_2/G=50$) coupled vdP oscillators. In all plots, blue curves correspond to $V/G=8$, while red curves correspond to $V/G=2$. The Fock basis  probability distributions for both these coupling strengths are shown in the insets (calculated for $\Delta=0$). The individual dissipation rate is $\kappa/G=0$.}
\label{Fig:revival_complete}
\end{figure}

To see an example of this, consider one quantum anharmonic vdP
oscillator to have a Kerr nonlinearity $K_1$, while the second vdP
oscillator is harmonic, i.e.~$K_2=0$. Deep in the quantum parameter regime, in which only the lowest three energy levels of each oscillator are populated, just three frequencies are relevant: The transition frequencies between the populated energy levels of the first oscillator, $\omega_1+K_1$ and $\omega_1+3K_1$, and the frequency of the second oscillator, $\omega_2$. The effective detuning between the two oscillators could therefore be minimized at two discrete values, $\tilde{\Delta}=\omega_2-\omega_1-K_1 = 0$ and $\tilde{\Delta}=\omega_2-\omega_1-3K_1 = 0$. At these values for which the effective detuning is minimized, we expect to see a revival of the oscillation amplitude. In Fig.~\ref{Fig:revival_complete}~(a), the blue curve depicts $\braket{a_1^\dagger a_1}=\braket{a_2^\dagger a_2}$ obtained by numerically simulating the master equation, Eq.~(\ref{master_equation}), for the example just described (the Fock level probability distribution is shown in the left inset). The peaks in the mean phonon number are clearly visible. The red curve in Fig.~\ref{Fig:revival_complete}~(a) depicts the peaks in $\braket{a_1^\dagger a_1}=\braket{a_2^\dagger a_2}$ for a smaller $V$, i.e.~for a parameter regime in which more Fock levels are populated (see right inset). Indeed, the peaks are seen in this case for $\Delta=(2n+1)K_1$, with $n$ being a nonnegative integer. The average oscillation amplitude squared, $\overline{|\alpha_1|^2}=\overline{|\alpha_2|^2}$, predicted by the semiclassical model, is shown in Fig.~\ref{Fig:revival_complete}~(b). As in Fig.~\ref{Fig:revival_complete}~(a), the blue and red curve correspond to $V/G=8$ and $V/G=2$, respectively. In both cases, only one peak is seen. This is expected, as the energy distribution is continuous in the semiclassical case. One can furthermore observe a mismatch in the peak location between the two cases. This is a classical effect, caused by the fact that the frequency of the nonlinear oscillator depends on the amplitude of oscillation, $\tilde{\omega}_1 = \omega_1 + 2K_1|\alpha_1|^2$. The peak appears for $\omega_2=\tilde{\omega}_1$, i.e.~for $\Delta=2K_1|\alpha_1|^2$. Smaller values of $V$ correspond to a larger amplitude of oscillation, and therefore the peak for $V/G=2$ appears to the right of the peak for $V/G=8$.

We now also consider the case in which both the vdP oscillators have an anharmonic energy spectrum. An example is shown in Fig.~\ref{Fig:revival_complete}~(c), in which the occupation number $\braket{a_1^\dagger a_1}=\braket{a_2^\dagger a_2}$ of both oscillators is plotted as a function of the detuning $\Delta$, for equal Kerr nonlinearities $K_1 =K_2\equiv K$. The blue curve corresponds to strong dissipative coupling $V$ for which only the first three low-lying Fock levels are populated (left inset), while the red curve corresponds to smaller $V$, for which more Fock levels have non-negligible population (right inset). We now expect phonon number peaks for $\Delta = 2nK$, with $n$ being an integer. These correspond to resonances between the transition frequencies of the two anharmonic oscillators, for the non-negligibly populated Fock states. The blue and red curve shown in Fig.~\ref{Fig:revival_complete}~(d) present the single peak which is predicted by the semiclassical model. Contrary to Fig.~\ref{Fig:revival_complete}~(b), and because both oscillators are nonlinear with $K_1=K_2$, both peaks appear at the same detuning $\Delta=0$.

\begin{figure}[t] 
\includegraphics[width=\columnwidth]{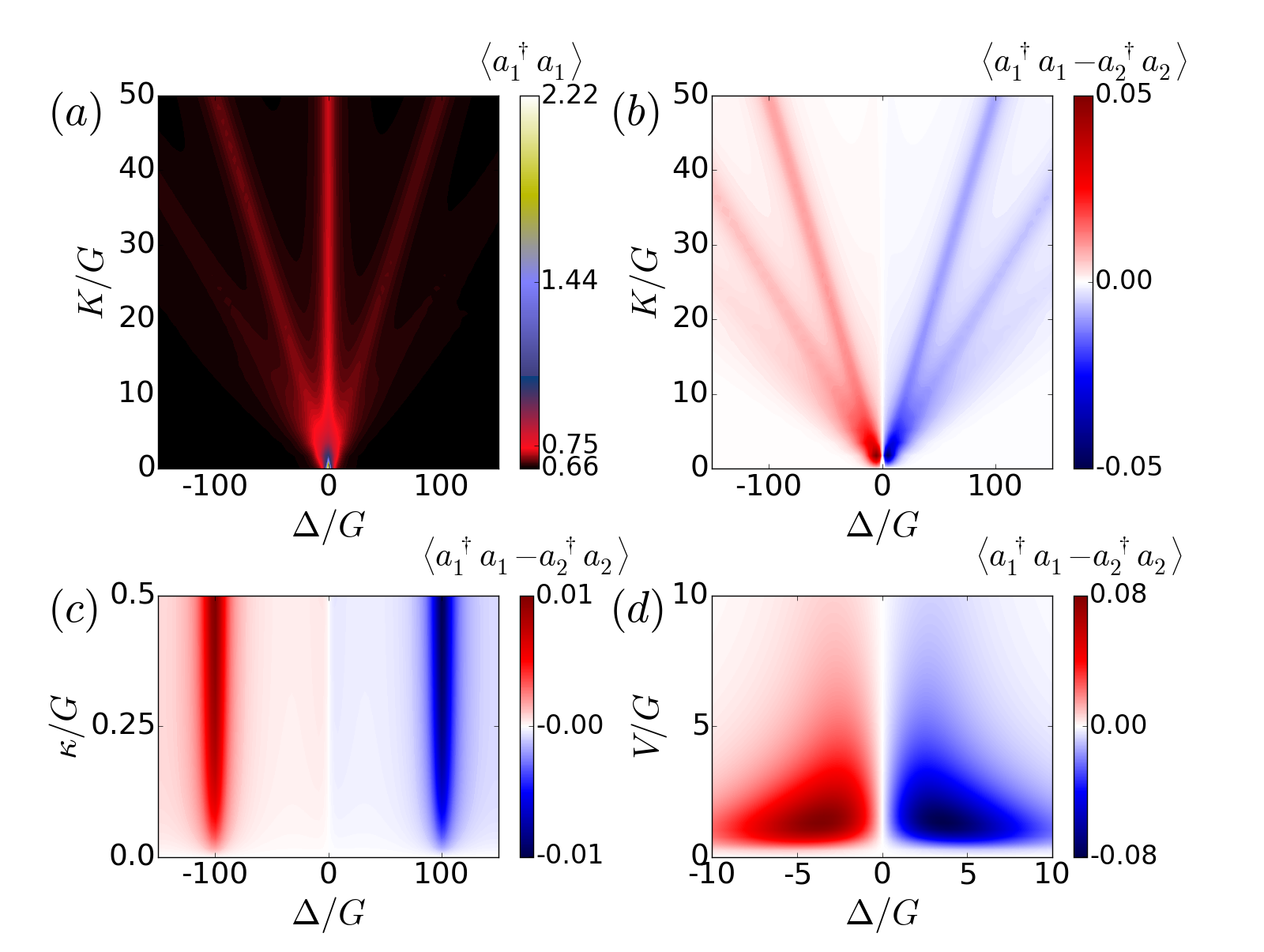}
\caption{(a) Average occupation number $\braket{a_1^\dagger a_1}$ as a function of $\Delta$ and $K$. Peaks in the occupation number at $K=\pm2\Delta$ are clearly visible. (b) Difference in occupation number between the two oscillators, $\braket{a_1^\dagger a_1 - a_2^\dagger a_2}$. The oscillation revival is more apparent in the $m$-th oscillator, if it involves its lowest frequency $\omega_m +K$. Other parameters are $(\kappa, V)/G=(0.25, 2)$ in both plots. (c) Average occupation number difference, $\braket{a_1^\dagger a_1 - a_2^\dagger a_2}$, as a function of $\Delta$ and $\kappa$. Other parameters are $(K, V)/G=(50, 2)$. The difference becomes more pronounced as $\kappa$ is increased. (d) $\braket{a_1^\dagger a_1 - a_2^\dagger a_2}$ for parameters corresponding to Fig.~\ref{Fig:harmonic_amplitudes}~(f), $(K, \kappa)/G=(1, 0.2)$.}
\label{Fig:Revival_Diff_combined}
\end{figure}

In the previously described examples, we set $\kappa/G=0$. The energy gain $G$ was balanced by the dissipative coupling $V$. This was possible because we have used large Kerr parameters $K_1$ or $K_2$, which therefore, as explained in Sec.~\ref{noise_induced_oscillation_death}, made the dissipative coupling more effective. For small values of $K_1$ and $K_2$, a finite value of $\kappa$ needs to be introduced in order to keep the system in the quantum parameter regime.

Figure~\ref{Fig:Revival_Diff_combined}(a) illustrates that in the absence of Kerr anharmonicity, i.e.~$K_1=K_2\equiv K = 0$, the two oscillators have a high phonon number only for $\Delta=0$. As $|\Delta|$ is increased, the oscillation-amplitude is strongly suppressed. For larger values of $K$, the oscillation amplitudes becomes much smaller, as the dissipative coupling is more effective. Still, for $\Delta =0$, we have a peak in the phonon number. As we increase $|\Delta|$, the phonon number decreases. But once $|\Delta|$ gets closer to the resonance condition $|\Delta|=2K$, the phonon number increases again. 

In Fig.~\ref{Fig:Revival_Diff_combined}~(a), a dissipation rate of
$\kappa/G = 0.25$ was chosen. It is needed to balance the energy gain
$G$ for the lower values of $K$. This value of $\kappa$ introduces a
small asymmetry between negative and positive detuning $\Delta$. In
Fig.~\ref{Fig:Revival_Diff_combined}~(b) the difference of the phonon
number between the two oscillators, $\braket{a_1^\dagger a_1 -
  a_2^\dagger a_2}$, is shown. We can see that for negative detunings,
the phonon number peaks are more pronounced for oscillator 1. For
positive detunings, the opposite is true. To understand this effect,
we need to consider the frequency resonances relevant to a
corresponding phonon number peak. For $\Delta>0$ ($\Delta<0$) and
$K>0$, the resonances involve the lowest possible transition frequency
of the second (first) oscillator, with higher transition frequencies
of the first (second) oscillator. As $\kappa$ is influencing energy
levels higher than the ground state, its effect is less detrimental on
the oscillator for which the lowest possible frequency is relevant. We
therefore expect that if the relations $K>0$ and $\kappa>0$ hold, such an asymmetry occurs. In Fig.~\ref{Fig:Revival_Diff_combined}~(c), the difference $\braket{a_1^\dagger a_1 - a_2^\dagger a_2}$ is plotted as a function of the detuning $\Delta$ and of $\kappa$ (other parameters are (V, K)/G=(2, 50)). It is indeed seen that $\braket{a_1^\dagger a_1 - a_2^\dagger a_2}=0$ for $\kappa=0$. As $\kappa$ is in increasing, so is the difference $\braket{a_1^\dagger a_1 - a_2^\dagger a_2}$. In Fig.~\ref{Fig:Revival_Diff_combined}~(d) we show the difference $\braket{a_1^\dagger a_1 - a_2^\dagger a_2}$ for parameters corresponding to Fig.~\ref{Fig:harmonic_amplitudes}~(f).

\section{Conclusions}   \label{Experimental_Realizations}

In this paper, we have studied theoretically the amplitude death phenomenon for two coupled anharmonic quantum vdP oscillators. We have shown that the anharmonicity leads to smaller oscillation amplitudes in semiclassical model and in the quantum description, an effect which we have shown to be the result of noise. Furthermore, we have found in the quantum description \emph{qualitative} differences as compared with the semiclassical model. Peaks in the mean phonon number of the oscillators are seen as a function of their detuning. They describe quantized amplitude death, and then oscillation revival. We have shown that these peaks correspond to discrete transition frequencies in the energy spectrum of the anharmonic vdP oscillators, and that they are therefore not seen  in a semiclassical model. To the best of our knowledge, this is the first time genuine quantum effects are discussed in the context of the amplitude death phenomenon.

Quantum vdP oscillators with a dissipative coupling can be engineered in a variety of systems. In trapped ion systems, one-phonon gain and two-phonon loss can be implemented using appropriately red- and blue-detuned drives \cite{Lee}.  The dissipative coupling can be implemented using various techniques \cite{Lee2, Schindler, Diehl}. For these systems, large Kerr nonlinearities $K_1$ and $K_2$ can be engineered \cite{Loerch, Zhao, Wang, Home}. In cavity optomechanical systems, quantum vdP oscillators can be realized using ``membrane-in-the-middle'' setups \cite{Walter, Thompson}. The two-phonon loss is obtained by placing the membrane at a node of the cavity field, and then driving the cavity with an appropriate red-detuned drive. The one-phonon gain is implemented via a coupling to another cavity mode, which is driven by an appropriate blue-detuned laser. A dissipative coupling of the form $\mathcal{D}[a_m-a_{\bar{m}}]$ between two such quantum vdP oscillators can be implemented using an additional cavity \cite{Walter}. Engineering large Kerr nonlinearities in optomechanical systems is extremely challenging and has not been demonstrated, but hybrid systems such as \cite{Rips, Rimberg} exploiting strong nonlinearities from auxiliary systems have been proposed. Realizing this experiment will demonstrate a new quantum effect in the amplitude dynamics of non-linear coupled oscillators.  


\section*{Acknowledgments}

This work was financially supported by the Swiss SNF and the NCCR Quantum Science and Technology. Numerical calculations were performed at sciCORE (http://scicore.unibas.ch/) scientific computing core facility at University of Basel.

\appendix

\section{Transforming the Fokker-Planck equation to a Langevin equation} \label{App:Langevin}

Following Ref.~\cite{Ishibashi}, we first rewrite the Fokker-Planck equation, Eq.~(\ref{FPE}), to Cartesian coordinates. Using $\alpha_m = x_m + iy_m$, we find
\begin{equation} \label{FPE_Cartesian}
\begin{aligned}
	\partial_t
	W(\boldsymbol{X})
&=
	\sum_{m=1}^2
	\left[
		-
		\left(
			\frac{\partial}{\partial x_m}
			\mu_{x_m}
		+
			\frac{\partial}{\partial y_m}
			\mu_{y_m}
		\right)
	+
		\frac{1}{2}
		\left(
			\frac{\partial^2}{\partial x^2_m }D_{x_m x_m}
		\right.
		\right.
\\
&
		\left.
		\left.
		+
			\frac{\partial^2}{\partial y^2_m }D_{y_m y_m}
		+
			\frac{\partial^2}{\partial x_m \partial x_{\bar{m}}}D_{x_m x_{\bar{m}}}
		+
			\frac{\partial^2}{\partial y_m \partial y_{\bar{m}}}D_{y_m y_{\bar{m}}}
		\right)
	\right]
W(\boldsymbol{X}),
\end{aligned}
\end{equation}
where $\boldsymbol{X}=(x_1, y_1, x_2, y_2)$, the drift vector $\boldsymbol{\mu}=(\mu_{x_1}, \mu_{y_1}, \mu_{x_2}, \mu_{y_2})$ is given by
\begin{equation}
\begin{aligned}
	\mu_{x_m}
=
	&\left[
		\omega_m
	+
		2K_m(x_m^2+y_m^2)
	\right]
	y_m
\\
&
+
	\left[
		\frac{G}{2} - \kappa(x_m^2+y_m^2-1) -\frac{V}{2}
	\right]
	x_m
+
	\frac{V}{2} x_{\bar{m}},
\end{aligned}
\end{equation}
\begin{equation}
\begin{aligned}
	\mu_{y_m}
=
	&-\left[
		\omega_m
	+
		2K_m(x_m^2+y_m^2)
	\right]
	x_m
\\
&
+
	\left[
		\frac{G}{2} - \kappa(x_m^2+y_m^2-1) -\frac{V}{2}
	\right]
	y_m
+
	\frac{V}{2} y_{\bar{m}},
\end{aligned}
\end{equation}
and the diffusion matrix is given by
\begin{equation}
\boldsymbol{D} =  \frac{1}{2} 
\left[\begin{array}{cccc}
    \nu_1 & 0 & -V/2 & 0	\\
    0 & \nu_1 & 0 & -V/2	\\
    -V/2 & 0 & \nu_2 & 0	\\
    0 & -V/2 & 0 & \nu_2
    \end{array}\right],
\end{equation}
where $\nu_m = G/2+\kappa\left[2(x_m^2+y_m^2)-1\right] +V/2$. 

The Langevin equation corresponding to Eq.~(\ref{FPE_Cartesian}) is
\begin{equation}
	\mathrm{d}
	\boldsymbol{X}
=
	\boldsymbol{\mu}
	\mathrm{d}t
+
	\boldsymbol{\sigma}
	\mathrm{d}\boldsymbol{W}_t,
\end{equation}
where $\mathrm{d}\boldsymbol{W}_t$ is the Wiener increment, and the noise strength is obtained via $\boldsymbol{\sigma}=\boldsymbol{U}\sqrt{\boldsymbol{D}'}\boldsymbol{U}^{-1}$, where $\boldsymbol{D}'=\boldsymbol{U}^{-1}\boldsymbol{D}\boldsymbol{U}$ is the diagonalized form of $\boldsymbol{D}$. As the Kerr nonlinearity does not appear in the diffusion matrix, the analytical derivation and exact expression for the noise term is identical to what is shown in Ref.~\cite{Ishibashi}.


\bibliographystyle{apsrev}
 \bibliography{Oscillation_Collapse}

\end{document}